\documentclass[twocolumn]{aa}
\usepackage{graphicx}
\usepackage{txfonts}
\begin{document}
   \title{Integral field near-infrared spectroscopy of II~Zw~40
   	\thanks{Based on observations obtained at the ESO VLT under program 078.B-0446.}}


   \author{L. Vanzi
          \inst{1,}  \inst{2},
          G. Cresci \inst{3}, E. Telles \inst{4,} \inst{5}
          \fnmsep\thanks{US Gemini Fellow}
          \and 
          J. Melnick \inst{1}.
          }

   \offprints{L. Vanzi}

   \institute{European Southern Observatory, Alonso de Cordoba 3107, Vitacura, Santiago - Chile\\
              \email{lvanzi@eso.org}
         \and
             Department of Electrical Engineering, Pontificia Universidad Catolica de Chile, Av. Vicu\~{n}a Mackenna 4860, Santiago - Chile
         \and
             Max-Planck-Institut f\"{u}r Extraterrestrische Physik, Postfach 1312, 85741 Garching - Germany 
          \and
             Observatorio Nacional, Rua Jos\'e Cristino 77, 20.921-400 Rio de Janeiro - RJ, Brazil
          \and
              Astronomy Department, University of Virginia, P.O. Box 400325,  Charlottesville, VA 22904-4325
              }

\date{Received ... ; accepted .... }

\abstract{We present integral field spectroscopy in the near-infrared of the nearby starburst galaxy
   II~Zw~40. Our new observations provide an unprecedented detailed view of the interstellar medium and
   star formation of this galaxy. The radiation emitted by the galaxy is dominated by a giant HII region, 
   which extends over an area of more than $400$ pc in size. A few clusters are present in this area, however 
   one in particular appears to be the main source of ionizing photons. We derive the properties of this object 
   and compare them with those of the 30 Doradus cluster in the Large magellanic cloud (LMC). We study the spatial 
   distribution and velocity field of different components of the inetrstellar medium (ISM), mostly through 
   the Bracket series lines, the molecular hydrogen spectrum, and [FeII]. We find that [FeII] and
   $H_2$ are mostly photon excited, but while the region emitting [FeII] is almost coincident with the giant 
   HII region observed in the lines of atomic H and He, the $H_2$ has a quite different distribution in space 
   and velocity. The age of the stellar population in the main cluster is such that no supernova (SN) should be 
   present yet so that the gas kinematics must be dominated by the young stars. We do not see, in the 
   starbursting region, any geometrical or dynamical structure that can be related to the large scale morphology 
   of the galaxy.
}
   
\authorrunning{Vanzi et al.}
   
\maketitle
%

\section{Introduction}
The starburst phenomenon, i.e., the occurrence of intense star formation in the central regions of some
galaxies, is one of the most interesting events taking place in the Universe. Through the detailed study of starburst galaxies in the nearby Universe, we can learn about star formation on a
galactic scale, a process quite distinct from what currently occurs in our Galaxy. In fact, unlike star formation observed in the Galaxy, starbursts produce a significant change in the chemical properties and in the energetic budget of the host galaxy. Because starburst galaxies have spectra similar to those of sub-millimiter galaxies and infrared luminous galaxies, they can be used to understand galaxies in the early universe, in particular, their formation and rapid evolution.
The bulk of galaxy formation is believed to have occurred
beyond redshifts of 1 (e.g., Hopkins \& Beacom 2006), and at those distances even the largest
objects are only a few arc-seconds in size (Cresci et al. 2006), preventing us from studying them in  detail. 
Blue dwarf galaxies (BDG) are objects with typically sub-solar abundances, which provide the opportunity to observe the starburst phenomenon in a relatively unevolved environment and with
less contamination by the underlying evolved stellar population than in larger galaxies, such as NGC 253 (Engelbracht et al. 1998). By studying BDGs we can access star formation events in a relatively pristine environment, but with details that cannot be achieved in the high redshift Universe. Therefore, these galaxies provide unique laboratories to study the mechanisms and the preferred modes of star formation during galaxy scale bursts.

At a distance of 10.5 Mpc (assuming a heliocentric velocity of 789 km/s de Vaucouleurs et al. 1991 and  $H_0$ = 75 km/s/Mpc), II~Zw~40 is one of the most interesting nearby starburst galaxies.
Despite being a small object, with a mass not reaching $10^9~M_{\odot}$ (Vanzi et al. 1996) and an absolute luminosity of $M_B \approx -18.2$, it hosts an extremely 
intense episode of star formation. It was first described by Searle \& Sargent (1972)
as a possible young galaxy because of its low-metallicity and because, at virtually any wavelength, the radiation emitted by this galaxy is dominated by
very young stars. The metallicity is 0.17 $Z_{\odot}$ (Gil de Paz et al. 2003).
The equivalent width of optical and near-infrared recombination lines is very large (Vanzi et al. 1996) and Wolf-Rayet features are detected in the optical spectrum (Vacca \& Conti 1992), both facts indicating a very young age. Spectroscopy in the mid-infrared shows evidence of high excitation from the high ratio of [SIV]/[NeII] (Wu et al. 2006).
The H$\alpha$ emission is mostly concentrated in a region of a few arcsec (Vanzi et al 1996). 
Beck et al. (2002) obtained a very large array (VLA)
radio image of this area and detected four compact sources within a region of about 0.5". The radio spectrum is thermal indicative of the presence of about 14000 O7 stars (Beck et al. 2002). 

The morphology of II~Zw~40 is highly irregular with two tails extending over a few tens of arcsecs which could be interpreted as parts of two different galaxies in the process of merging, and is often considered responsible for triggering the intense central starburst (Baldwin et al. 1982).  HI imaging  shows  the presence of a flat extended  HI structure that could be a highly-inclined disk, or a pair of tidal-tails of neutral hydrogen emerging from the galaxy and extending over more than 6 arcmin. Van Zee et al. (1998) find a strong radial velocity gradient along these tails with a clear velocity reversal in the SE tip reminiscent of the tidal tails observed in the prototypical merger NGC~7252 where the velocity reversal is caused by material falling back toward the dynamical center (Hibbard \& Mihos, 1995).  Notice, however, that as stressed by van Zee et al. (1998),  the colliding galaxies in NGC~7252 are much more massive than II~Zw~40, for which no merger models have yet been calculated. Therefore, while the HI observations are highly suggestive of a merger of two roughly similar disk galaxies, the evidence is not yet conclusive, and we must be very cautious when interpreting kinematical data of the central regions of II~Zw~40.

II~Zw~40 is one of those BDGs with direct evidence for molecular gas though its geometry is not known. It was detected in CO(1-0) by Arnault et al. (1988), Gondhalekar et al. (1998) and Tacconi \& Young (1987). In the latter work, the authors derived an $H_2$ mass of $2.87\times10^7 M_{\odot}$. Sage et al. (1992) also detected II~Zw~40 in CO(1-0) and CO(2-1), finding a centroid velocity of 770 km/s and a  $H_2$ mass lower limit of $5\times10^6 M_{\odot}$. II~Zw~40 was not detected in CO(3-2) by Meier et al. (2001).

\begin{figure*}[t]
\centering
\includegraphics[width=17cm, angle=0]{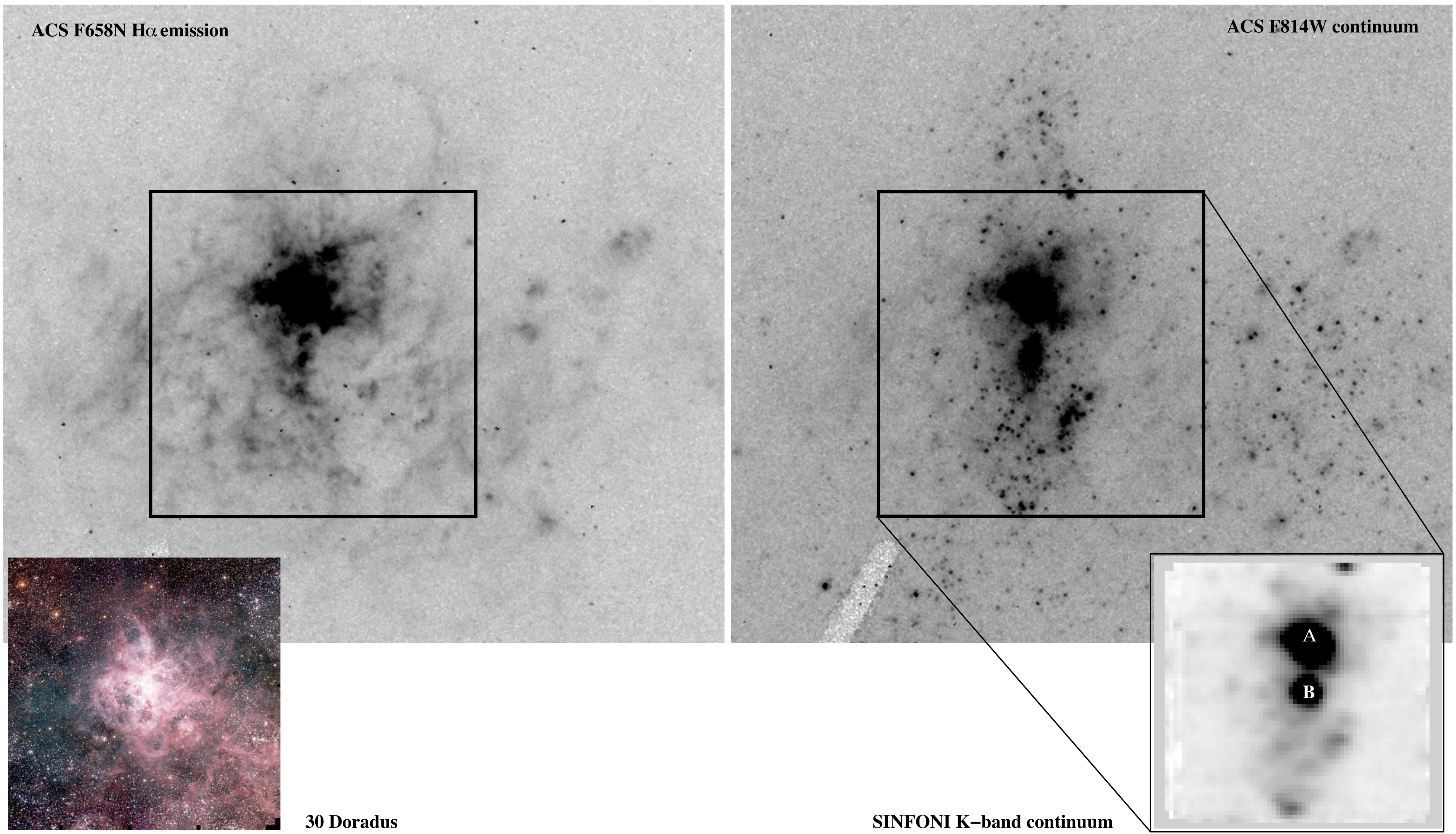}
      \caption{Archive H$\alpha$ image of II~Zw~40, obtained by ACS on board the HST (left panel). The central box indicates the field covered by our SINFONI observations, FOV = 8 \arcsec. The insert in the bottom left shows the 30 Doradus complex, image from WFI at the LaSilla 2.2 m telescope, ESO press photo 14a/02. Archive F814W image of II~Zw~40, obtained by ACS on board the HST (right panel). In the lower right insert, continuum image in K band obtained with SINFONI, the field of view is 8 arcsec. North is up, east to the left.}
\label{ACS}
\end{figure*}

\begin{figure}
\centering
\includegraphics[width=7cm, angle=0]{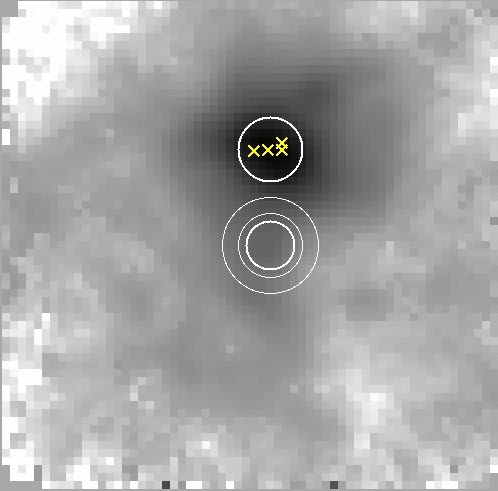}
      \caption{Close-up view of II~Zw~40 in H$\alpha$ displayed in logaritmic scale. The extraction apertures of the spectra around sources A and B are indicated, the anulus used to evaluate the background emission around source B is also indicated. The positions of the VLA radio sources of Beck et al. (2002) are marked by crosses.}
\label{Ha}
\end{figure}

In Fig.~\ref{ACS}, the ACS H$\alpha$ image of II~Zw~40 is shown. This image shows essentially all the features that characterize the largest giant HII regions in nearby galaxies: a very bright core and a number of loops and filaments apparently emanating from the core that resemble closely what is observed in the 30 Doradus complex (c.f.  bottom-right insert of Fig.~\ref{ACS}).
Figure ~\ref{ACS} also shows a continuum ACS image at 814nm where a number of bright point-like objects can be seen that must be bright supergiants and star clusters in the galaxy.  The brightest continuum source coincides with the H$\alpha$ peak and clearly shows the super-star-cluster that ionizes the giant HII region. Interestingly, while the complex morphology observed in the central 15pc by Beck et al. (2002) with the VLA, is well matched by the H$\alpha$ image, the optical continuum is unresolved in the ACS images and coincides in position with the central VLA source (Fig. \ref{Ha}).  At the position of the other three radio sources, the optical continuum is diffuse, indicating that these strong thermal radio sources may be very young massive clusters still deeply embedded in their placental clouds (Fig. \ref{Ha}).  The diffuse optical emission may be free-free radiation from the hot gas.  We conclude that the bright HII region that dominates the emission-line morphology of II~Zw~40 is ionized by a single super-star cluster $\sim 0.1''$ in diameter, corresponding to about 5pc, i.e., about the same size of the 30 Doradus cluster (Selman et al., 1999).

In this paper, we present an investigation of the interstellar gas and star formation in the central region of II~Zw~40, using the IR adaptive-optics assisted-integral field spectrometer on the very large telescope (VLT) SINFONI.   With a spectral resolution of almost 4000 in K-band and almost 3000 in the H-band, SINFONI enables a detailed  study of the 3D properties of ionized and molecular gas within a radius of 200 pc from the central ionizing sources with unprecedented angular resolution resolution and with enough spectral power  to investigate the kinematics of the gas.

The paper is organized as follows. In Sect. 2 we describe the observations and the analysis of the 
data. In Sects. 3 and 4 we present the results of the compact sources detected and the ISM, respectively. In Sect. 5, we present a comprehensive picture of the galaxy, as it can be derived from our study. Finally, in Sect. 6 we summarize our conclusions.


\section{Observations and Analysis}

II~Zw~40 was observed with SINFONI (Eisenhauer et al. 2003) at the VLT in October and November 2006 on three different nights. Two spectra were obtained in each of the H and K grisms providing a full coverage of these bands and a
spectral resolution of about 3000 and 4000, respectively.The total integration time on-source was 40 minutes in each band. We used the  0.25 "/pix scale which 
provides integral field spectroscopy over a  field of view of $8 arcsec \times 8$ arcsec.
Adaptive optics (AO) correction was provided by the system MACAO, the brightest central region
of the galaxy was used to close the loop. The observations were obtained under excellent seeing conditions (typically 0.6 \arcsec), but because of the spatial extension and limited brightness of the  AO reference source only partial correction could be obtained reaching a Strehl ratio of about 15\%.

The spectra were reduced using the ESO SINFONI pipeline version 1.3.
The correction for the telluric features and the flux calibration were obtained by dividing by the
spectra of solar type stars observed in the same configuration as the main target. The spectral
features introduced in this way were corrected using a spectrum of the sun, following the
procedure described by Maiolino et al. (1996). The flux calibration of spectra obtained on different nights was consistent within 10\%.

We used the tool QFitsView \footnote{http://www.mpe.mpg.de/\~{}ott/QFitsView} to extract images centered on selected emission lines and 1-D spectra at well defined locations in the field observed. In particular, we obtained single line images for the brightest lines detected, Br$ \gamma$ and Br 11; He2.06 and He1.70;
the line of [FeII] at 1.64 $\mu m$ and the lines of $H_2$ at 2.195, 2.212, 2.222, 2.224, 2.240,
and 2.242 $\mu m$. We extracted the images with a band equivalent to a spectral resolution
of about 1100. The continuum was subtracted from a nearby spectral region free of any
emission or absorption features.
We extracted 1-D spectra at two positions centered on the two main sources visible in the
field. 
Finally, we extracted radial velocity maps and velocity dispersion maps centered on Br$\gamma$, 
[FeII]1.64, and $H_2$2.12. Further, 1-D spectra were extracted in the region of bright $H_2$ emission,
details are given in Sec. 5.4.

\section{The compact sources}
The continuum images extracted from the SINFONI data cubes show two prominent sources and a
number of much fainter ones, lower right panel of Fig. \ref{ACS}. The brightest source, which we will indicate as source A, is located to the north and slightly elongated in the NE-SW direction.  It encompasses the four thermal radio sources discussed above. The fainter source to the south is round and we will indicate it as source B.

\begin{figure*}
\centering
\includegraphics[width=12cm, angle=0]{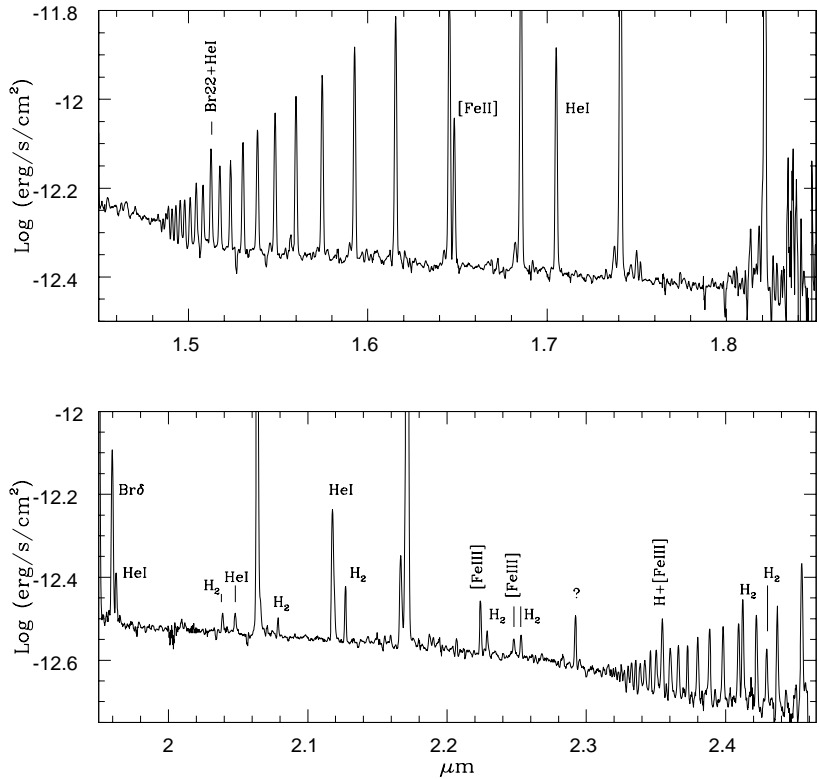}
      \caption{Near-infrared spectrum of source A extracted with a circular aperture of 1" radius. Besides a few obvious lines such as Br$\gamma$ at 2.17 $\mu m$ or HeI at 2.06 $\mu m$, a large number from the Bracket series are detected in the H spectrum, and from the Pfund series beyond 2.3 $\mu m$. Less obvious lines are explicitly indicated. }
\label{SpcA}
\end{figure*}

We extracted the spectrum of source A with a circular aperture of 4 pixels in radius, corresponding to
1" or about 50 pc. A large number of emission lines are detected in the spectrum as shown in Fig. \ref{SpcA}. The main lines detected are identified in Table 1 and their observed fluxes are listed. The errors in the absolute values are mostly due to the flux calibration and must be considered to be on the order of 10\% (see Sec. 2).
The equivalent width (EQW) of Br$\gamma$ is 440 \AA, which is close to the maximum value reached by a cluster of young stars formed in a single burst episode. According to Starburst 99 (SB99 - Leitherer et al. 1999), for a single burst of solar metallicity, Br$\gamma$ has an early maximum of about 500 \AA~, it drops below 440 \AA~ only after 3 Myr and has a sharp drop between 6 and 8 Myr. For metallicity $Z_{\odot} /5$, the situation is mostly unchanged, but the drop is slightly smoother occurring between 5 and 10 Myr. This sets an upper limit of $\sim3$~Myr to the age of the cluster. The average extinction measured in this area from the ratio Br$\gamma$ / Br11 is $A_V = 6.3 $, (see Sect. 5.3). The dereddened Br$\gamma$ flux measured in this region is $2.8\times10^{-14} erg/s/cm^2$, which gives a luminosity of $3.5\times10^{38} erg/s$ for a distance of 10.5~Mpc.

\begin{figure*}
\centering
\includegraphics[width=13cm, angle=0]{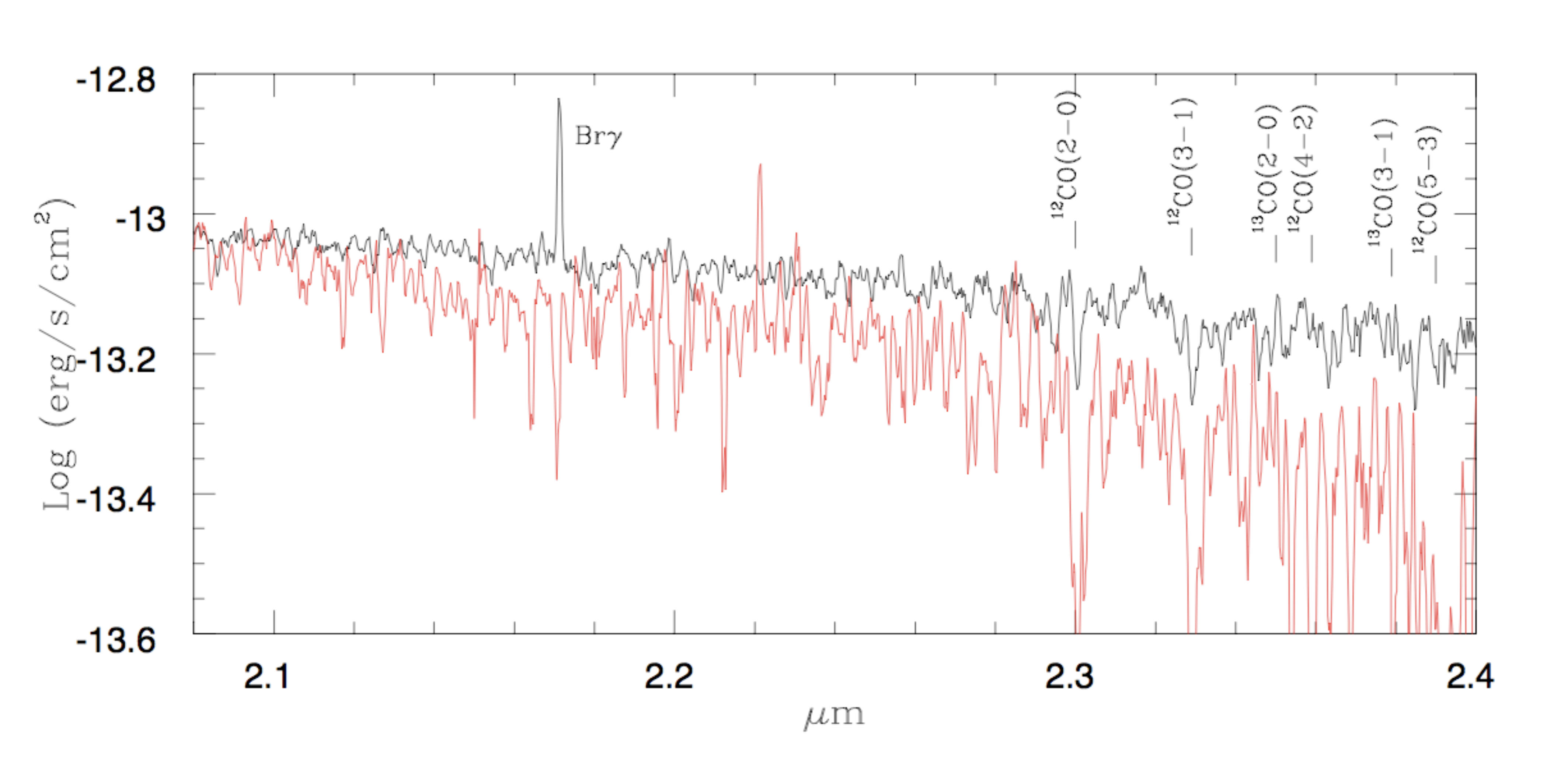}
      \caption{Near-infrared spectrum of source B in the K band dark line. The spectrum was extracted with a circular aperture of 0.75" radius and subtracted from the spectrum of the surrounding area. The CO stellar band head are indicated. The red line shows the combined spectrum of the faint
      sources in the field of SINFONI: the spectrum is scaled up by a factor of about 10.}
\label{SpcB}
\end{figure*}

\begin{table*}
\caption{Main emission lines detected in the spectrum of II~Zw~40 A, the spectrum was extracted with a
circular aperture of 4 pixel or 1.0 \arcsec radius. The observed fluxes are given in units of $10^{-16} erg/s/cm^2/\mu m$ and the absolute errors are about 10\%.  }
\begin{center}
\begin{tabular}{ccc|ccc}
\hline
line        & $\lambda_{obs.} (\mu m)$ &    Flux    &   line        & $\lambda_{obs.} (\mu m)$ &    Flux      \\
\hline
 Br 30     &    1.485    &    1.31        &    Br$\delta$   &   1.944   &    64.5    \\
 Br 29    &    1.487    &    1.05        &    He I + \verb1[FeII]?1 &   1.954   &    6.94   \\
 Br 28    &    1.489    &    1.25        &    H2 1-0 S (3) &   1.957   &    1.14   \\
 Br 27    &    1.491    &    1.55        &    H2 1-0 S (2) &   2.034   &    0.35  \\
 Br 26    &    1.494    &    1.52        &    He I               &   2.042   &    0.33  \\
 Br 25    &    1.497    &    1.67        &    He I               &   2.058   &    34.7  \\
 Br 24 + ?  &    1.500    &    2.25    &    He I               &   2.061   &    0.15  \\
 Br 23    &    1.504    &    2.19        &    H2 2-1 S (3) &   2.073   &    0.43    \\
 Br 22 + HeI   &    1.508    &    3.89 & He I               &   2.113   &    5.27    \\
 Br 21    &    1.513    &    3.08        &    H2 1-0 S (1) &   2.122   &    1.35   \\
 Br 20    &    1.519    &    3.18        &    H2 2-1 S (2) &   2.154  &    0.17   \\
 Br 19    &    1.526    &    4.08        &    He I               &   2.162  &    3.24   \\
 Br 18    &    1.534    &    5.04        &    Br$\gamma$&  2.165  &  103.4   \\
    ?        &    1.545 (obs.)    &    0.31        &    He I                &  2.184  &    0.16   \\
 Br 17    &    1.544    &    5.84        &    \verb1[FeIII]1              &  2.218   &    1.35  \\
 Br 16    &    1.556    &    6.85        &    H2 1-0 S (0)  &  2.223   &   0.57   \\
 Br 15    &    1.570    &    8.45        &    \verb1[FeIII]1              &  2.242   &   0.36   \\
 Br 14    &    1.588    &  10.0          &    H2 2-1 S (1)  &  2.248   &   0.48   \\
 Br 13    &    1.611    &  13.3          &    \verb1[SeIV]?1      &  2.287   &   1.15   \\
 Br 12    &    1.641    &  17.0          &    Pf 35              & 2.325   &   0.41   \\
\verb1[FeII]1 &    1.643    &    5.90          &    Pf 34              & 2.328   &   0.53   \\
\verb1[FeII]1 &    1.664    &    0.15          &    Pf 33             & 2.332   &   0.55  \\
\verb1[FeII]?1   &    1.682 (obs.)   &   1.40           &    Pf 32            &  2.335   &   0.61  \\
 Br 11    &    1.681    &   23.0         &   Pf 31              &   2.339   &   0.69  \\
 He I      &    1.700    &   10.5         &   Pf 30              &   2.343   &   0.82  \\
 He I      &    1.732    &   1.20         &   Pf 29 + \verb1[FeIII]1 &   2.348   &   1.68  \\
 Br 10    &    1.736    &   32.2         &   Pf 28              &   2.354   &   1.01  \\
\verb1[FeII]1 + HeI  &   1.745  &  0.73        &   Pf 27              &   2.360   &    1.00 \\
 He I      &    1.748    &   0.40         &   Pf 26              &   2.366   &    1.02 \\
\verb1[FeII]1      &    1.809    &  1.45           &   Pf 25              &   2.374   &   1.22  \\
 He I      &    1.813    &  1.66          &   Pf 24              &   2.382   &    1.56 \\
   ?         &    1.820 (obs.)   &  3.45          &   Pf 23              &   2.392   &    1.64  \\
  Br 9  &     1.817    &  48.6          &   Pf 22              &  2.403    &    1.68 \\ 
              &                   &                    &   H2 1-0 Q (1) &  2.406    &    2.44  \\
              &                   &                    &   Pf 21 + H2 1-0 Q (2)  &  2.413  &  1.97  \\
              &                   &                    &   H2 1-0 Q (3) &  2.424   &     1.12   \\
              &                   &                    &   Pf 20              &  2.431   &     2.11  \\
              &                   &                    &   H2 1-0 Q(4)  &  2.437   &     0.25  \\
              &                  &                     &   ?                     &  2.451 (obs.)   &     0.40  \\
              &                  &                     &   Pf 19              &  2.448   &     3.96 \\
\hline
\end{tabular}
\label{lines}
\end{center}
\end{table*}

The Br$\gamma$ flux within a larger circular aperture of 3" in radius is $3.4\times 10^{-14} erg/s/cm^2$. The average extinction in this larger area is about $A_V\sim4$ magnitudes, giving a dereddened flux of $5.1\times10^{-14} erg/s/cm^2$ to be compared with the value of $4.5\times10^{-14} erg/s/cm^2$ measured within the same aperture by Vanzi et al. (1996). 
The total Br$\gamma$ flux over most of the emission line area, integrated on a circle of 15" in diameter and corrected for extinction is $5.8\times10^{-14} erg/s/cm^2$. We used the parameters of early-type stars of Panagia (1973) to derive that this flux requires about 9000 equivalent O7 stars.
However, the radio continuum observations require 14.000 O7 stars  (Beck et al., 2002) consistent with the fact that three of the VLA radio sources do not have optical counterparts in the ACS continuum imaging and are faint in H$\alpha$, as discussed above.  These sources could be very young and still deeply-embedded clusters.
Assuming, therefore, that a single source is responsible for the ionization of this entire region, we can use the
total Br$\gamma$ luminosity to derive the mass of source A.  For a single stellar population, 3~Myr old and a Kroupa IMF,  we obtain a mass of $1.7\times10^6 M_{\odot}$ using SB99.

Source B is significantly fainter both in Br$\gamma$ and continuum. 
We extracted a 1-D spectrum centered on source B with a circular aperture of 3 pixels in radius. As
stated in Sect. 1 it is likely that only the brightest source provides ionizing photons. For
this reason, we subtracted from the spectrum of source B, the spectrum of the large nebula surrounding source A measured on an annular region around source B, and assumed that it is mostly due to source A. The equivalent width of Br$\gamma$, after the nebular subtraction, is 10 \AA~ and the flux is $7.7\times10^{-17} erg/s/cm^2$. Correcting for the average extinction that on source B is Av = 4.9, and comparing these numbers with SB99, we obtain an age between 6 and 7 Myr and a total
mass of about 1.3 $10^5 M_{\odot}$ for a single starburst episode.  At an age of 6-7~Myr, the eqw of the emission lines drops very rapidly and descends basically to zero beyond 10 Myr. A very  interesting aspect of source B is the detection of the CO band-heads at $2.29 \mu m$ (see Fig. \ref{SpcB}), a signature of the presence of red supergiants. Using the definition of Doyon et al (1994) to measure a spectroscopic CO index, we derived a value of 0.05, which according to SB99 based on the same definition, corresponds to an age of 7.5~Myr, in very good agreement with the age determination from Br$\gamma$. Although [FeII] 1.64 is very weak for source B, the ratio [FeII]/Br$\gamma$ is almost 4 times larger than for source A, indicating a possible contribution from SN, which would also indicate an age in agreement with the one estimated above (see Sect. 5.2 for a more quantitative analysis).

A gaussian fit over the continuum image of sources A and B gives FWHM of $1.6 \arcsec \times 1.2$ \arcsec and $1\arcsec$ diameter, respectively. We do not have precise knowledge of our PSF, but as the seeing during the observations was about $0.6 \arcsec$, we can be confident that both sources are resolved. Deconvolving with a $0.6''$ worst-case PSF, we obtain a size of $75 pc \times 50 pc$ for source A and 40 pc for source B. The ACS images, however, show that both sources are very compact with a FWHM at the limit of the resolution of HST, which is about 0.1 \arcsec, or about 5 pc. The much larger sizes measured in the IR, both in $Br\gamma$ and continuum, must be due to the extent of the HII region. In fact, according to Vanzi et al. (1996), a large fraction of the near-IR continuum is due to free-free emission. 

The other sources are too faint and their extracted spectra do not show any feature, either in emission or in absorption. We made the arbitrary assumption that these sources are somehow alike and combined their spectra to increase the signal-to-noise ratio. It was possible to detect the 2.29 CO band head in absorption on this combined spectrum with an index of 0.20, which according to SB99  indicates ages between 8 and about 40 Myr. In conclusion, the starburst region of II~Zw~40 is one of those cases powered by a single young massive cluster (e.g., Vanzi \& Sauvage, 2004).

\section{The interstellar medium of II~Zw~40}
Our 3-D data cubes provide essential information about the ISM in II~Zw~40. Spectroscopy in the near-IR
gives access to different phases of the ISM: the ionized gas through the recombination lines of H and He, but also of other species such as Fe; the warm molecular gas through the $H_2$ emission; and the dust through the ratios of well-modeled lines as tracers of the extinction. 

\subsection{Atomic Hydrogen and Helium}

\begin{figure*}
\centering
\includegraphics[width=21cm, angle=0]{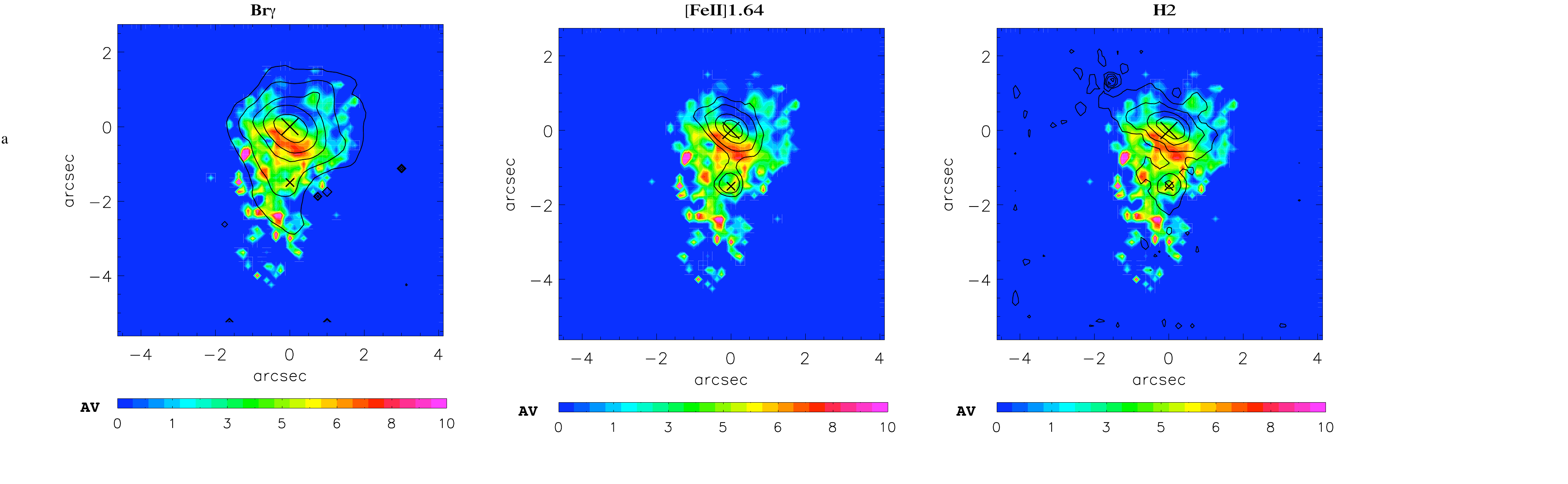}
      \caption{Comparison of the geometry of the emission lines (contours) and the spatial distribution of the extinction, derived from the Br$\gamma$ / Br11 ratio (colorscale). The contours of $Br\gamma$, [FeII]1.64, and $H_2$ are plotted on the left, central, and right panel respectively, they are 3, 8, 20, 40 and 100~$^{-14} erg/s/cm^2$ for Br$\gamma$; 0.6, 1, 2, 3, 8 and 10~$^{-14} erg/s/cm^2$ for [FeII]; and 0.1, 0.3, 0.6, 1, 2~$^{-14} erg/s/cm^2$ for $H_2$. The positions of the brightest continuum sources A and B are indicated by crosses. }
\label{estinzione}
\end{figure*}

The photoionized gas is mostly observed through the recombination lines of Hydrogen and Helium. Several of these lines are detected in our spectra. We extracted an image centered on the $Br\gamma$ line with a bandwidth of 20 \AA. The emission peaks on source A and shows an elongation similar to the one observed in the continuum  in the NE - SW direction with FWHM of 5.6 and 7.6 pixels, or 1.4 and 1.9 arcsec. In the left panel of Fig. \ref{estinzione},
we present the contours of Br$\gamma$ over-plotted to the extinction map of the galaxy ( see Sect. 5.3). The Br$\gamma$ peak coincides with the peak of the continuum, indicated with a cross in the image, the position of source B is also marked by a cross. The $Br\gamma$ emission extends over an area that covers our field of view almost entirely, but our image is not deep enough to fully evidence the structures observed in H$\alpha$ by HST. 
No other bright $Br\gamma$ source is present besides source A.


$Br\delta$ is detected at the blue edge of the K band in a region of poor atmospheric transmission while the entire Bracket series is detected in the H band starting from Br 9. The spatial distribution of the He lines is very similar to the one given by the H lines. Vanzi et al (1996) used the ratios He 1.70 and 2.11 $\mu m$ to Br10 and Br$\gamma$, respectively, as indicators of the stellar temperature. On source A we measure $0.33\pm0.01$ and $0.051\pm0.005$ for these two ratios, respectively. The first value is in agreement with the measurement of Vanzi et al. (1996) and their theoretical prediction of 0.31, the
second one lies on the high side of the predicted value 0.040. This is not completely unusual; Armand et al. (1996) observed similarly a value for 2.11/$Br\gamma$ on the high side of the prediction in one of their ultracompact HII regions. We extracted a He1.70/Br10 map and found that the ratio is constant over an area of about 2 \arcsec diameter around source A, outside this area the line ratio cannot be measured anymore.

Although the galaxy is classified as WR based on the detection of  He WR emission features in the optical, we do not detect any of the corresponding features expected in the near-IR (Lumsden et al. 1994).

\subsection{Iron}
\begin{figure}
\centering
\includegraphics[width=12 cm, angle=0]{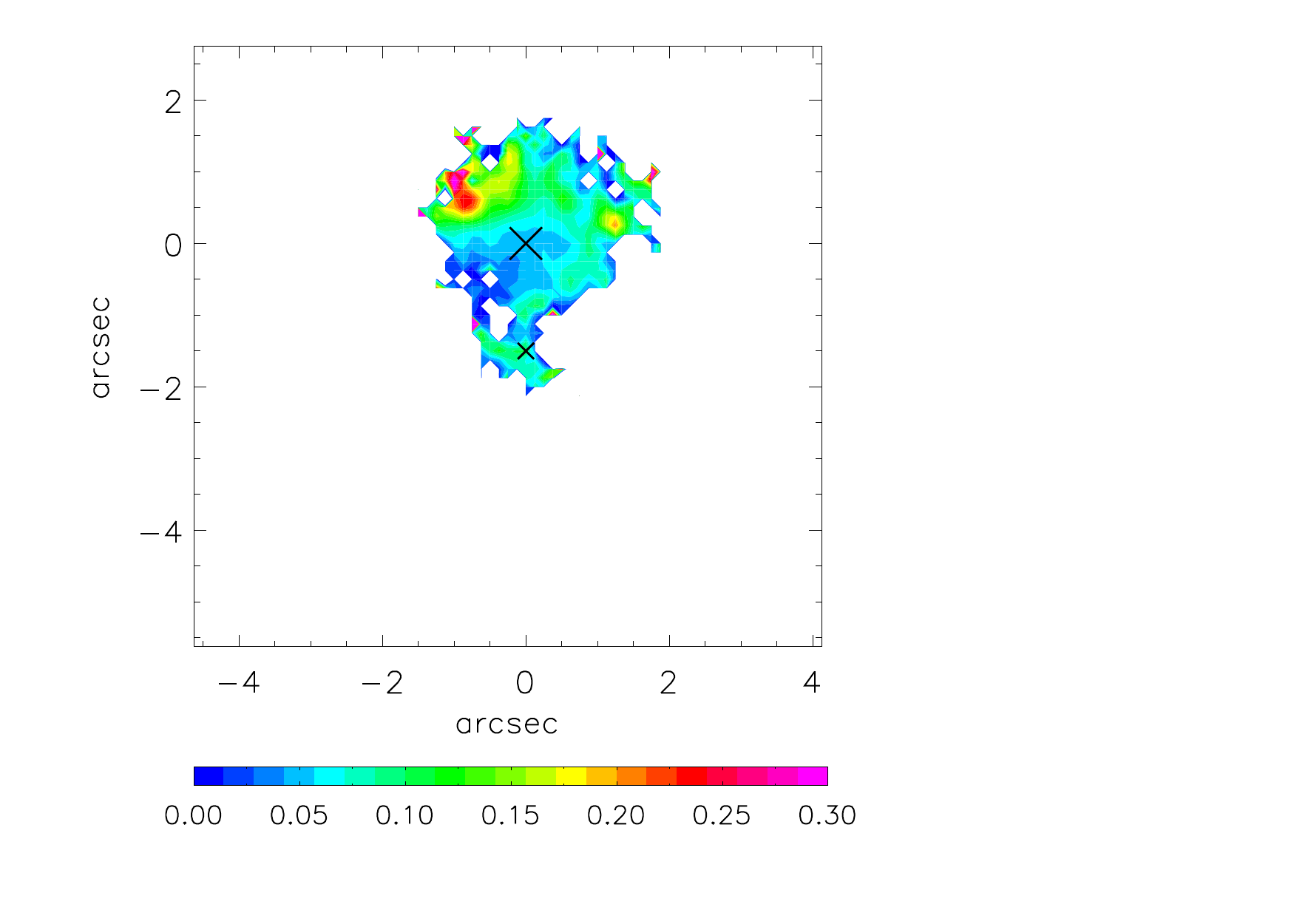}
      \caption{Map of the [FeII]/Br$\gamma$ ratio.The value of the ratio is mostly in agreement with the typical values for HII regions with the exception of few isolated areas where it could be enhanced by the presence of a limited numbers of SN remnants. The positions of sources A and B are marked by crosses.}
\label{iron}
\end{figure}

The emission of the forbidden lines of [FeII] occurs in partially-ionized regions. Such regions can be found at the edge of HII regions where, however, they are thin. Much more extended  regions of partial ionization are produced by shocks (Mouri et al 2000). For this reason the [FeII] emission lines has proved to be an excellent tracer of SN remnants.  The brightest Fe line present in our spectra is  [FeII]1.64 $\mu m$; its ratio to $Br\gamma$ is  0.06 in the Orion nebula as compared to more than 30 in SN remnants (Moorwood \& Oliva 1988; Oliva et al. 1990). We extracted an image centered on the line [FeII]1.64, and plotted the contours in the center
panel of Fig. \ref{estinzione}. The comparison with the contours of Br$\gamma$ shows that the centroids are almost coincident, with [FeII] being slightly shifted to the NE  and slightly more elongated. We derived an extinction free [FeII]/ Br$\gamma$ image, using Br 11 and rescaling to Br$\gamma$ (Fig. \ref{iron}). The ratio [FeII]/ Br$\gamma$ is not uniform, but has a value of about 0.06 at the position of source A with a gradient that increases from SE to NW from 0.02 to about 0.12. It is  about 0.1 on a U-shaped area pointing to NW and it is maximum in an area  $1.2''$ to the NE of source A, with values up to 0.2).

Values around 0.06 are basically compatible with pure photoionization and do not require the presence of supernovae, consistent with the extremely young age of source A.  We must therefore, assume that most of the [FeII] flux is generated in the giant HII region associated with source A. The higher values observed toward
the NE may indicate the position of an older cluster, or even a supernova remnant whose [FeII]/Br$\gamma$ ratio gets diluted by the Br$\gamma$ emission of the giant HII region. Another [FeII] bright compact source is detected to the NW of source A. From Fig. \ref{estinzione}, it is already evident that source B is more prominent in [FeII] than it is in Br$\gamma$, the ratio [FeII]/Br$\gamma$ on this source after removing the nebular contribution from source A, as described in Sect. 4, is 0.25, indicating a possible contribution from SN remnants. This is roughly consistent with the age estimate of source B derived in Sect. 4.  The [FeII] flux measured on source B is $1.83\times10^{-17} erg/s/cm^2$, which gives a luminosity of $5.0\times 10^{35} erg/s$,  after correcting for extinction.

We can use SB99 to derive the number of SN remnants expected in source B for the mass and age derived. We find a supernova rate of  $1.3\times10^{-4}$ SN/yr, somehow lower than the value derived by Vanzi et al. (1996), which, however, should be considered as an upper limit. Assuming a [FeII] emitting phase of about $10^4$ yr, we expect that between 1 and 2 SN remnants could be [FeII] bright in source B. Vanzi \& Rieke (1997) found a [FeII]1.64$\mu m$ average luminosity of $8\times10^{36} erg/s$ per SN remnant. Alonso-Herrero et al. (2003) measure values between $3\times10^{36}$ and $2.2\times10^{38} erg/s$ and other authors (Oliva et al. 1989, Lumdsen \& Puxley 1995) give lower values. In conclusion, the [FeII]1.64 luminosity that we measure is low, but consistent with the expectations, especially considering the low number statistics of  SN remnants in source B.
We must also consider that while the SN rate is relatively metallicity independent, the [FeII] luminosity depends on the metal content, so that our low value could, in part, be the result of the
sub-solar abundances in II~Zw~40.

Lines of [FeIII] are detected at rest wavelength 2.219, 2.243, and 2.348 $\mu m$, the latter blended with
H Pfund 29. The [FeIII] line at 2.145 $\mu m$ is not detected. We compared the ratios of these lines with those calculated by Bautista \& Pradhan (1998) as a function of the density for $T_e = 9000 K$. We obtained I(2.24)/I(2.22)=0.21$\pm$ 0.10, I(2.35)/I(2.24)=3.5$\pm$1.5, and I(2.22)/I(2.15)$>$ 13.5 $\pm$ 3.7. All ratios are indicative of low densities,
$10^2~cm^{-3}$ or less,  all are significantly smaller than the values observed in Orion.

\subsection{Dust}
We used images centered on $Br\gamma$ and Br11 to derive an extinction map of the galaxy. We assumed an intrinsic ratio Br11 / $Br\gamma$ = 0.25,  (Hummer \& Storey 1987 for T=$10^4$K and n=$10^2 cm^{-3}$). Jaffe et al. (1978) estimated the foreground Galactic extinction in the direction of II~Zw~40 to be $A_V = 1.2$. Our extinction map, presented in colorscale in Fig. \ref{estinzione}, clearly shows that the extinction is not uniformly distributed. An area of higher extinction is located to the south of source A whereas an area of lower extinction is observed to the north. Both areas show an elongation similar to the one observed for source A. An asymmetry in the dust distribution is, in fact,  often observed  in galactic HII regions where young stars clear the ISM in preferential directions. In particular, in II~Zw~40 an extinction ridge is located between source A and source B.

To compare our IR measurements with the optical, we used the work of Kong \& Cheng (2002), who measured $A_V = 3.07$ with a slit of 3 \arcsec, whilst using a similar aperture we obtain $A_V = 3.37$. We derive the dust content from the extinction assuming a ratio N(HI) / Av = $5.8~ 10^{21} cm^{-2}/mag$ (Bohlin et al. 1978), and a gas-to-dust mass ratio of 1000 (Lisenfeld \& Ferrara 1998). The average extinction over a region of 610 pix, equivalent to 38 arcsec$^2$ or about $10^5 pc^2$, is 3.78 mag. From 
this value we derive a dust mass of $1.6~ 10^4  M_{\odot}$.  Hunt et al. (2005) find a much higher value, $1.8\times10^5 M_{\odot}$  from their fit of the spectral energy distribution of the galaxy. Even larger values are found by Galliano et al. (2005), who get $0.4-1.1\times10^6 M_{\odot}$ also from a fit of the SED. In both cases, their results are sensitive to all the dust that emits in the IR, whereas we can only detect the dust that absorbs the background radiation along the line of sight. Thus, the discrepancy with our results may indicate the presence of an extended component of warm dust to which we are insensitive, or the presence of a dense core of dust, molecular gas, and young stars heavily obscured and not visible at 2 $\mu m$, also indicated by the large ionizing flux observed at radio wavelengths discussed in Sect. 4. 

%

\subsection{Molecular Hydrogen}
We detected several lines of $H_2$ in our spectra. We extracted images centered on the brightest lines and combined them in a single image to increase the S/N. The contours of this image are plotted in the right panel of Fig. \ref{estinzione}. The difference with $Br\gamma$ and [FeII] is conspicuous.  The centroid of the emission is shifted to the northeast and it is obviously off source A. In addition, the $H_2$ emitting region is significantly more extended and elongated in the northeast direction, reaching an area where we see no emission from $Br\gamma$ and [FeII] at all. 

The $H_2$ lines in the near infrared are predominantly excited through two mechanisms: absorption of UV photons (fluorescence) and collisions (thermal). The two mechanisms mostly excite different roto-vibrational levels so we can discriminate between the two processes using the ratios of a few lines and comparing, for example, with the models calculated by Black \& van Dishoek (1987). Lines generated by vibrational levels with $v \ge 2$ are virtually absent in thermally excited spectra, while they are relatively strong in fluorescent spectra. High vibrational level lines are abundant in the H spectrum, but they are faint and difficult to detect because of the presence of strong sky lines. The 2-1 lines, on the other hand, can be detected in the K spectrum, as for instance, 2-1S(1) at 2.25 $\mu m$ and 2-1S(0) at 2.36 $\mu m$. 

We extracted three 1-D spectra covering different areas of the $H_2$ emitting region, mainly to the NE (2.7\arcsec north, 4.5\arcsec east of the pick of Br$\gamma$), to the SW (basically coincident with the pick of Br$\gamma$), and in the center of the $H_2$ emitting region, 1.2\arcsec north, 1\arcsec east of the Br$\gamma$ pick) (C). In Table~2 we list the ratios of the lines detected to 1-0 S(1) at 2.12 $\mu m$. Following Engelbracht et al. (1998) in Table 2 we compare the ratios of the lines detected to 1-0S(1) at 2.12 $\mu$ m with the models of Black \& van Dishoek (1987). The H band lines and the 2-1S lines in the K band spectrum are expected to be very weak for thermal excitation. However, several H band lines are detected in region C and all the  2-1S K band lines are detected in the three regions, with the exception of 2-1 S(4) at 2.01 $\mu m$, which lies in a spectral region of poor atmospheric transmission.  This strongly suggests fluorescence excitation process. 
We obtained maps of the most significant line ratios and verified that there is no significant deviation from the behavior sampled by the ratios reported in Table 2, in particular, we do not detect any region where shocks significantly contribute to the excitation of $H_2$.

The detection of photon excited $H_2$ in region NE, where basically no $Br\gamma$ emission is detected, and the striking difference of geometry between $H_2$, and $Br\gamma$, [FeII] are quite remarkable and underline the power of integral spectroscopy for the study of starburst galaxies. 
\begin{table}
\caption{Comparison of the observed ratios of $H_2$ to $H_2$ (1-0) S(1) at 2.12 $\mu m$ in 3 different regions of the galaxy (see text) with fluorescent and thermal excitation models by Black \& van Dishoek (1987). The observed line ratio are in agreement with a fluorescent excitation model.}
\begin{center}
\begin{tabular}{ccccccc}
\hline
line         &    $\lambda (\mu m)$  &  NE   &      C      &     SW     &     fl.     &     th.      \\
\hline
5-3Q(1)  &  1.493  &  -         &     -        &       -        &  0.43  & 1.0 $10^{-4}$    \\
4-2O(3)  &  1.510  &  -         &  0.23     &       -        & 0.42  & 7.0 $10^{-4}$    \\
6-4Q(1)  &  1.601  &  -         &  0.20     &       -        & 0.33  & 1.3 $10^{-4}$    \\
5-3O(3)  &  1.613  &  -         &  0.25     &   0.17     & 0.38  &  9 $10^{-5}$      \\
6-4O(3)  &  1.733  &  -         &  0.23     &   0.12     & 0.31  &  1 $10^{-5}$      \\
\hline
1-0S(3)  &  1.958 &   0.69   &  1.14     &   0.98     &  0.67  &  0.93   \\
2-1S(4)  &  2.004 &   -          &   -          &    -           &  0.12  &  0.02   \\
1-0S(2)  &  2.034 &   0.43   &  0.41     &   0.32     &  0.50  &  0.38   \\
2-1S(3)  &  2.073 &   0.24   &  0.38     &   0.32     &  0.35  &  0.08   \\
2-1S(2)  &  2.154 &   0.19   &  0.21     &   0.18     &  0.28  &  0.04   \\
1-0S(0)  &  2.223 &   0.46   &  0.53     &   0.37     &  0.46  &  0.21   \\
2-1S(1)  &  2.248 &   0.41   &  0.45     &   0.34     &  0.56  &  0.08   \\
2-1S(0)  &  2.356 &   0.15   &  0.27     &   0.16     &  0.26  &  0.02   \\
3-2S(1)  &  2.386 &    -         &   -          &   0.25     &  0.29  &   5.8 $10^{-3}$  \\
1-0Q(1)  &  2.407 &   0.77   &  1.18    &   1.62     &   0.99  &  0.70   \\
1-0Q(2)  &  2.413 &   -          &  0.23    &   -           &   0.51  &  0.23    \\
1-0Q(3)  &  2.424 &   0.46   &  0.69    &   0.75     &   0.70  &  0.70    \\
1-0Q(4)  &  2.437 &   -          &  0.14    &   0.15     &   0.28  &  0.21   \\
\hline
Br$\gamma$   & 2.167  &     -     &   11.7    &     66.6  &    &      \\
\hline
\end{tabular}
\label{h2}
\end{center}
\end{table}
We detect quite a strong line at rest wavelength 2.287 $\mu m$, which could correspond to the transition (3,2) S(2) of $H_2$. According to the fluorescent model of Black \& van Dishoeck (1987), however, the ratio of this line to 1-0 S(1) should be 0.14, while we measure 0.87. One explanation of such a large difference of flux may be the line identified as due to [SeIV] by Dinerstein (2001) and observed by Sterling et al. (2007) in a number of planetary nebulae. No line is found instead at the location of the other line observed by these authors at 2.199 $\mu m$ and attributed to [KrIII].

\subsection{Velocity field}
\begin{figure*}
\centering
\includegraphics[width=21cm, angle=0]{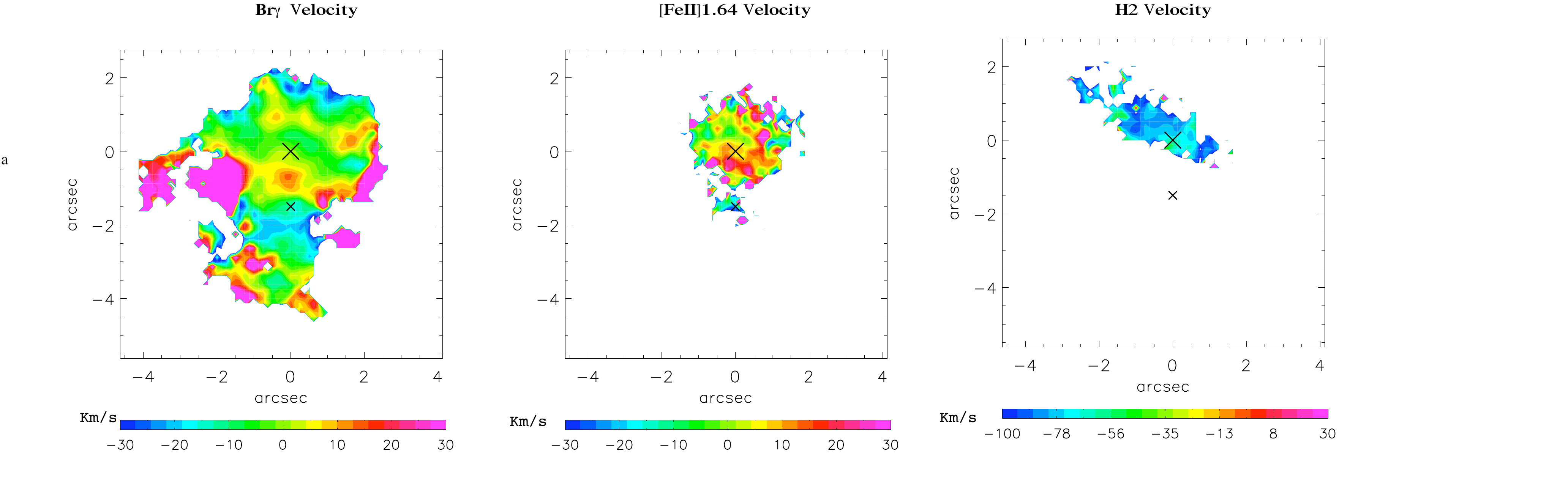}
      \caption{Radial velocity maps of $Br\gamma$ (left), [FeII]1.64 (center) and $H_2$ (right). The positions of the brightest continuum sources A and B are marked by crosses. The zero point of the velocity refers to the Br$\gamma$ emission on source A. The $H_2$ cloud shows an average offset of about 90 km/s with respect to the other components.}
\label{vel}
\end{figure*}

\begin{figure*}
\centering
\includegraphics[width=18cm, angle=0]{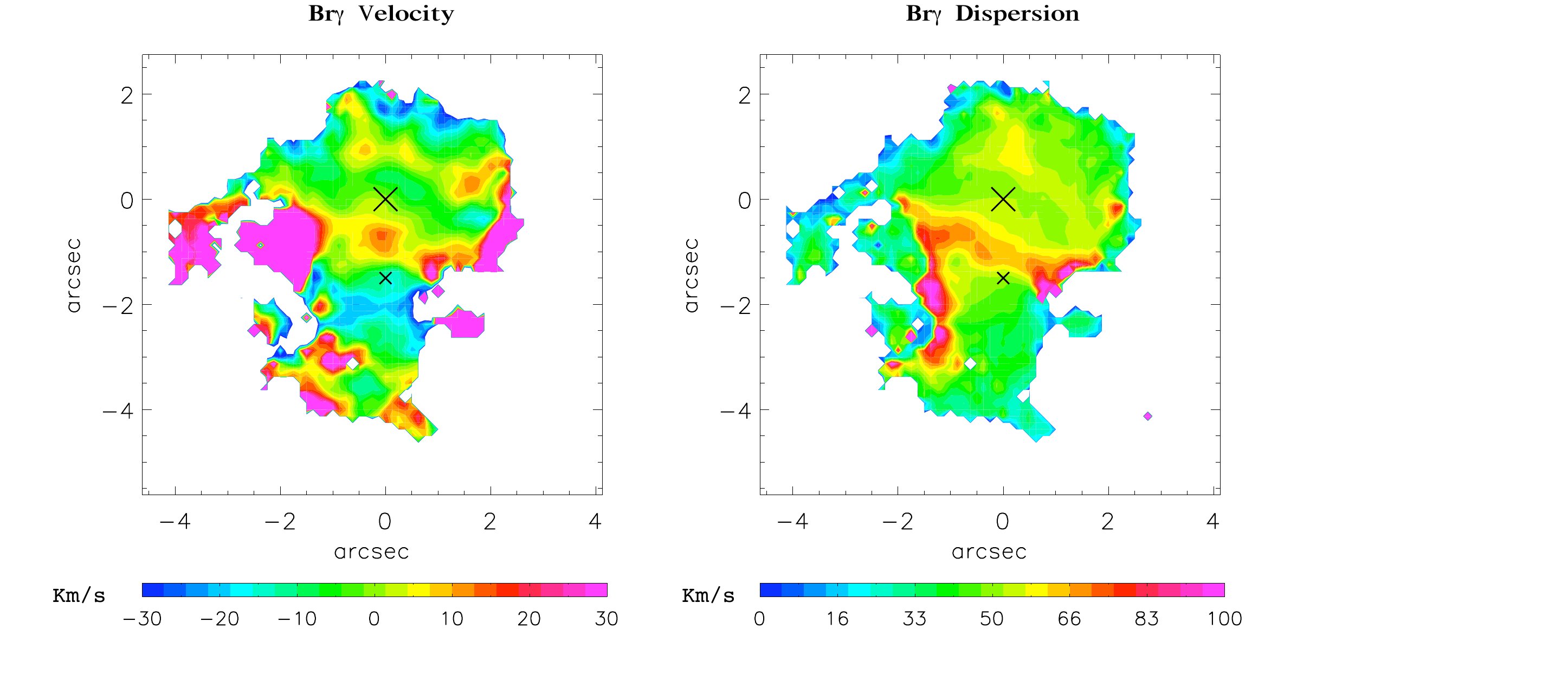}
      \caption{Radial velocity (left) and velocity dispersion (right) maps of Br$\gamma$. The position of sources A and B are marked by crosses.}
\label{sigma}
\end{figure*}

We derived radial velocity and velocity dispersion maps for the relevant lines by fitting gaussians to the observed lines profiles. These maps, corrected for instrumental broadening, are shown in Figs.~\ref{vel} and \ref{sigma}, where the positions coincident with the continuum peak of the two main clusters are marked by a cross. The  radial velocity of  Br$\gamma$  at the position of the peak of source A defines the zero point of the velocity scale. The errors on the radial velocities depends on the S/N and range from 1 to 10 km/s in Br$\gamma$, from 1.5 to 35 km/s in [FeII], from 2.4 to 45 km/s in $H_2$, areas with too large errors are not covered in the maps.

Some features can be seen in the radial velocity map of Br$\gamma$,  mostly showing structures and orientations similar to those observed in the extinction map and as the elongation of source A. The nature and origin of these structures are difficult to interpret. Their shape and orientation do not seem to be connected to the optical tails of the galaxy, which are oriented toward south and southeast. They correspond to  regions where the line profiles are double or multiple indicative of wind-driven expanding bubbles (Bordalo et al. 2007). The HI disk observed by van Zee et al. (1998) is oriented in the SE-NW direction, almost parallel to one of the optical tails. This radio structure, which is more than 6 arcmin across, has a total gradient of more than 100 km/s with velocities lower than 760 km/s toward NW and more than 860 km/s in the SE extreme, suggesting  that the radio structure is dominated by rotation.  On the other hand, the velocity structures that we observe do not show comparably large velocity gradients, and their geometry does not appear related to the 21cm structure. Thus, the dynamics of the start-bursting region is different and does not show a clear connection with the large scale dynamics of the galaxy. 

The maps of [FeII] and $H_2$ cover a smaller region and are mostly flat. They show an intriguing difference in velocity among the lines: while the Br$\gamma$ average velocity is close to zero, [FeII] is redshifted by about 10-15 km/s and, most remarkably, $H_2$ is blueshifted by almost 90 km/s. So not only is the $H_2$ cloud spatially distinct  from the giant HII region at the center of II~Zw~40, but it also has a very different velocity field.

We derived a velocity dispersion map for Br$\gamma$ only. The velocity dispersion, deconvolved for the instrumental profile of 
$\sigma_{inst}=38$~km/s, appears quite
uniform over the region observed, with values close to $\sigma\sim30$~km/s, which are quite typical of HII galaxies (Terlevich et al. 1992).  We also observe areas
with high velocity dispersion coincident in some cases with regions of high radial velocity where the profiles typically show multiple components, indicative of wind-driven expanding bubbles. SNRs do not seem likely to play a relevant role in shaping the dynamics  of the region for the reasons discussed above. Both the radial velocity and velocity dispersion maps of Br$\gamma$ are consistent with the  H$\alpha$ observations of Bordalo et al. (2007), although our maps have a higher angular resolution (but cover a smaller area). They observe that the regions showing the highest variation of radial velocity and velocity dispersion are those with the lowest flux, while the brightest regions consistently show the values of velocity and dispersion observed in the integrated spectrum. They argue that the kinematics of the brightest regions is mostly driven by gravity while the faint regions would be mostly affected by winds.

\section{Discussion}

\subsection{The central giant HII region of II~Zw~40}

In order to discuss the properties of the central giant HII region in II~Zw~40 (II~Zw~40A), we compare them to those of the prototypical giant HII region 30 Doradus in the LMC, which has been extensively studied by numerous authors (see Bosch et al. 2001 and references therein). Table~\ref{comparison} summarizes the relevant properties of the two objects.  

\begin{table}
\caption{Properties of 30 Doradus versus II~Zw~40. (1) Selman et al. (1999); (2) Walborn (1991); (3) Kennicutt (1984); (4) Werner et al. (1978); (5) Peimbert et al. (2005); (6) Chu and Kennicutt (1994);  (7) Beck et al. (2002); (8) Vanzi et al. (1996); (9) Melnick et al. (1988).}
\begin{center}
\begin{tabular}{lcc}
\hline
                               & 30 Doradus        &    II~Zw~40   \\
 \hline
Cluster size (pc)           & 20   (1)                    &    25 (7)   \\
Nebular size (pc)         & 200 (2)                 &  400      \\
$\rm Log L H{\alpha} (erg/s)$       & 40.2 (3)       & 41.0 (8)      \\
$\rm Log L_{FIR} (L_{\odot})$              & 7.6 (4)         & 9.1 (8)      \\
SFR surface density ($M_{\odot} yr^{-1} kpc^{-2}$) & 380 & 1600 \\
12+log(O/H)                 & 8.59 (5)             & 8.13 (9) \\
$\rm \sigma (km s^{-1})$ & 22 (6)           & 35.2  (9) \\
age (Myrs)                    & 2-3 (1)               & 3 \\
\hline
\end{tabular}
\label{comparison}
\end{center}
\end{table}

In spite of being ionized by starburst clusters of apparently similar linear sizes, II~Zw~40A is significantly more energetic than 30 Doradus. 
The luminosity of II~Zw~40 (both nebular and stellar) is about one order of magnitude larger than 30 Doradus, indicating a much larger star formation rate per unit area. The gas in II~Zw~40 is significantly more turbulent indicating that the ionizing cluster of II~Zw~40 is considerably more massive according to the relation between nebular velocity dispersion and luminosity (Melnick et al. 1987). As discussed in Sect.~4, there is a discrepancy of a factor of 1.5 between the ionization flux derived from the nebular lines (in our case Br$\gamma$), and the one derived from the VLA radio continuum observations. We used the H$\alpha$ luminosity to derive the star formation rate following Kennicutt (1998) and we adopted the size of the VLA source in Table~\ref{comparison} to calculate the star formation surface density of Table \ref{comparison}. Thus, the resulting specific star formation rate of 1600~$M_{\odot} yr^{-1} kpc^{-2}$ is realistic and is not very different to the values derived for high-redshift starburst galaxies (e.g. Blain et al. 2002 and references therein). The star formation rate turns out to be of about 1~$M_{\odot}/yr$ in II~Zw~40 and 0.15 ~$M_{\odot}/yr$ in 30 Doradus, so that although the star formation occurs in very similar areas, it is significantly more efficient in II~Zw~40. In spite of this difference, the morphologies of both HII regions are remarkably similar, which is reassuring since the ionizing clusters have similar ages and in particular have not reached the phase where the gas dynamics would be dominated, or at least strongly influenced, by SN explosions.  This indicates that, for a given age and metallicity, one should be able to scale the properties of local HII galaxies to interpret the observations of young galaxies at large redshifts. We should caution however, that, as shown above, the central HII region in II~Zw~40 does not partake in the overall dynamics of the parent galaxy, which seems to be dominated by rotation. This is probably not the case for very luminous high redshift star-forming galaxies where the observations indicate the presence of multiple giant HII regions  distributed over a much larger area that collectively reflect the galaxian dynamics.

\subsection{Other components}

Besides the central giant HII region - powered by one super-star cluster - there are several other fainter sources visible in the central part of II~Zw~40. They are likely to be clusters with older ages and possibly lower masses. The most prominent of them, named source B throughout this paper, has a mass of $1.3\times10^5 M_{\odot}$ and an age of about 7 Myr.  While the stellar population of the brightest cluster (source A) is very young showing no evidence of SN explosions, a few SN remnants may be present in source B. 

Around source A the ISM is entirely excited by the ionizing photons emanating from the central cluster,  with no evidence for different excitation mechanisms  apart from a possible contribution from SNe that, however, is limited to few compact areas visible to the northeast and northwest of source A in Fig. \ref{iron}. The extinction is irregularly distributed. The ridge in the extinction map, located between source A and source B, corresponds to a region of high radial velocity and high velocity dispersion, probably due to compression of the ISM by stellar winds.

One of the most striking findings of our observations is the presence of a large cloud of molecular hydrogen whose geometry and kinematics appear to be completely disconnected from the giant HII region that dominates the galaxy. This cloud has a differential velocity of about 90 km/s with respect to the rest of the galaxy, and while the HI observations of van Zee et al. (1998) provide a marginal indication of neutral components with these extreme velocities in the central region of the galaxy,  their spatial resolution is by far not enough to associate these components with our $H_{2}$ map. The projected size of this cloud is about $4\times1$ arcsec or  $200\times50$ pc, which for  a density of $10^2 cm^{-3}$ gives a crude estimate for the mass of a few $10^6 M_{\odot}$ comparable to the masses of giant molecular clouds in our Galaxy.  While this cloud is dynamically distinct from the central region of the cluster, it  must be sufficiently close to the ionizing source to be fluorescently excited and to glow. It may, therefore, be falling in, or escaping from, the center of the galaxy. This would be consistent with the merger scenario, but high resolution HI and CO observations would be required to investigate this further, especially because we could check whether molecular gas is being fed onto the central regions of the galaxy as a consequence of the merger.  Pockets of $H_{2}$ are also observed within the 30 Doradus HII region where one would a-priori expect all remaining molecular material to be evaporated by the strong radiation field ionizing the nebula (Rubio et al. 1998). Unfortunately, however, these regions have only been observed photometrically and their radial velocities remain unknown.  

As the $H_2$ traces the dense warm molecular gas, we conclude that in II~Zw~40, $H_2$ must be distributed in large clouds, of which we detect one, which does not necessarily share the dynamics of the bulk of the galaxy. The fact that basically no $H_2$ emission is detected at the radial velocity, corresponding to the giant HII region itself, indicates that the UV radiation field must be strong enough in this area to destroy the $H_2$. It is interesting that the $H_2$ cloud has a structure and orientation similar to the extinction pattern, to some velocity structures, and to the elongation of source A, as if all this was actually part of a unique picture resulting, for instance, from a compression wave that eventually triggered the star formation. 

To our knowledge, it is the first time that this picture is observed. Similar observations have been obtained mostly of AGN or starburst/AGN composit objects. The results in these cases are not uniform. Davis et al. (2006) find a complex morphology for the $H_2$ in the Sy NGC 3227, in particular with a peak off the galaxy nucleus. Davis et al. (2004) find instead similar kinematics and geometry for all ISM components in the ULIRG Mrk 231. Finally, Zuther et al. (2007) observed the composite object Mrk 609 where the $H_2$ geometry follows the continuum, and there are hints for the [FeII] and $H_2$ to be dynamically decoupled.

\section{Conclusions}

 \begin{enumerate}
      \item We obtained integral field spectroscopy in the near-infrared with AO correction of the central star forming region of II~Zw~40. This region was shown to be dominated by one giant HII region powered by a single young super-massive cluster. This cluster has a mass of $1.7\times 10^6 M_{\odot}$ and an age of  3 Myr at most, as derived from the Br$\gamma$ luminosity and equivalent width, respectively. All other compact sources detected are less massive and older.
      \item II~Zw~40 appears to be one of those starburst galaxies that, when observed with high enough
      resolution, appears to be powered by one single young massive cluster. We compare the characteristics of this object with those of 30 Doradus in the LMC and find that the starbursting region of II~Zw~40 could be regarded as a scaled up version of it.  
      \item A comparison of published VLA radio continuum maps with HST/ACS images indicates the presence of two or three very young massive clusters in the center of the II~Zw~40 deeply enshrouded in dust.
      \item The ISM is mostly photo-excited. Even the lines of $H_2$ and [FeII] do not show signs of shocks. Consistent with the very young age of the region there must be very few, if any, SN in the area, certainly none would be present in the main cluster.
      \item We detected a giant cloud of $H_2$, which is not related to the giant HII region, having a 
      different morphology and dynamic, but that is still photo-excited by the radiation field emanating from the main cluster.
      \item The dynamics of the ISM in the star forming region does not seem to follow, or  be determined by,
    a larger scale pattern, and since SNe are virtually absent must be fully dominated by the star formation.
      \item We detect structures in the ISM that are consistently oriented and we speculate that they are
      the effects of a common cause, such as, for instance, a compression of the ISM that eventually triggered the
      star formation.
\end{enumerate}

\begin{acknowledgements}
Based on observations made with the NASA/ESA Hubble Space Telescope, obtained from the data archive at the Space Telescope Institute. STScI is operated by the association of Universities for Research in Astronomy, Inc. under the NASA contract  NAS 5-26555. This research has made use of the NASA/IPAC Extragalactic Database (NED), which is operated by the Jet Propulsion Laboratory, California Institute of Technology, under contract with the National Aeronautics and Space Administration. 

The authors thank Chris Lidman for useful and enlightening discussions and Alberto Rodriguez-Ardila for reading and commenting the manuscript. LV and GC thank the Observatorio Nacional de Rio de Janeiro for support during their visit. LV acknowledges support from the ESO Director General Discretionary Funds.
      Finally, we thank the anonymous referee for valuable comments and suggestions that substantially improved the final version of this paper.
\end{acknowledgements}

\end{document}